\documentclass[12pt]{article}
\def\dom{\mbox{dom}}
\def\deg{\mbox{deg}}
\def\s#1#2{\mbox{$\Sigma_{#2}^{#1}$}}
\def\p#1#2{\mbox{$\Pi_{#2}^{#1}$}}
\def\Union{\bigcup}

\def\I#1#2{\mbox{$\bigcap_{#2}{#1}_{#2}$}}
\def\Seq#1#2{\mbox{$\{{#1}_{#2}\}$}}

\def\lth#1{|#1|}
\def\join{\oplus}

\def\into{\longrightarrow}
\def\Teq{\equiv_{T}}
\def\Tle{<_{T}}
\def\Tleq{\leq_{T}}

\def\LL{{\cal{L}}}
\def\scriptS{{\cal{S}}}
\def\TT{{\cal{T}}}
\def\UU{{\cal{U}}}
\def\VV{{\cal{V}}}
\def\DD{{\cal{D}}}
\def\Eone{\mbox{DTIME}(2^{\mbox{\scriptsize\rm linear}})}
\def\Etwo{\mbox{DTIME}(2^{\mbox{\scriptsize\rm polynomial}})}

\def\strings{\{0,1\}^*}
\def\cylinders{\{0,1,\perp\}^*}
\def\languages{\{0,1\}^{\infty}}
\def\defined#1{||#1||}
\def\NP{\mbox{\rm NP}}
\def\P{\mbox{\rm P}}

\def\Iff{\Longleftrightarrow}
\def\NN{\mbox{\rm I\hspace{-.12em}N}}

\def\Pr{\mbox{\rm\bf Pr}}
\def\Ext{\mbox{\rm Ext}}

\newtheorem{theorem}{Theorem}[section]
\newtheorem{lemma}[theorem]{Lemma}
\newtheorem{claim}[theorem]{Claim}
\newtheorem{definition}[theorem]{Definition}

\newenvironment{proof}{\begin{trivlist}\item[]\hspace{\parindent}%
{\em Proof.}}{$\bullet$\end{trivlist}}

\newenvironment{proofof}[1]{\begin{trivlist}\item[]\hspace{\parindent}%
{\em #1}}{$\bullet$\end{trivlist}}

\def\alt#1#2#3#4{\left\{\begin{array}{ll} #1\  & \mbox{ if } #2\\
#3\  & \mbox{ if } #4\end{array}\right.}

\newenvironment{enumerate.roman}%
{\begin{enumerate}}%
{\end{enumerate}}

\newenvironment{enumerate.alph}%
{\begin{enumerate}}%
{\end{enumerate}}

\title{Independence Properties of Algorithmically Random Sequences}

\author{Steven M. Kautz \\
Department of Mathematics \\
Randolph-Macon Woman's College \\
2500 Rivermont Avenue \\
Lynchburg, VA 24503}

\date{April 5, 1995}

\begin{document}

\maketitle

\begin{abstract}
A bounded Kolmogorov-Loveland selection rule is an
adaptive strategy for recursively selecting a subsequence
of an infinite binary sequence; such a subsequence
may be interpreted as the query sequence of a time-bounded
Turing machine.  In this paper we show that if $A$ is
an algorithmically random sequence, $A_0$ is selected
from $A$ via a bounded Kolmogorov-Loveland selection rule,
and $A_1$ denotes the sequence of nonselected bits of $A$,
then $A_1$ is {\em independent\/} of $A_0$; that is, 
$A_1$ is algorithmically random relative to $A_0$.
This result has been used by Kautz and Miltersen 
\cite{Kautz-Miltersen94}
to show that relative to a random oracle, NP does 
not have $p$-measure zero 
(in the sense of Lutz \cite{Lutz92}]).

\end{abstract}

\section{Introduction}
Any plausible characterization of what it means for an infinite
sequence to be {\em random\/} is likely to incorporate the common notion
that one random sequence should not contain any information about
any other random sequence, or further, that
observing {\em part\/} of a random sequence should provide no
information about any other part of the same sequence.
These ideas can be found in the intuitive definition of
random sequences proposed by R. von Mises in the 1920's
\cite{vonMises28}:  what makes
a sequence $A$ random, he suggested, is that
\begin{enumerate.roman}
\item
{\em limiting frequencies\/} exist, i.e., 
\[
\lim_{n \rightarrow \infty} \frac{\mbox{\# of 1's in $A[0..n-1]$}}{n} = p
\]
for some number $p$
(where $A[0..n-1]$ denotes the first $n$ bits of $A$), and
\item
given any admissible rule for selecting a subsequence $A_0$
from $A$, the limiting frequency in $A_0$ is the same value $p$.
\end{enumerate.roman}

The exact meaning of
the term ``admissible'' is problematic and controversial,
at least in part because there
certainly exists a selection rule of the form
``select bit $A[n]$ iff $A[n] = 1$.''  Church \cite{Church40}
proposed limiting the ``admissible'' rules to be recursive functions
which given an initial segment $A[0..n-1]$ produce either a zero
or one, according to whether the $n$th bit is to be selected.
A sequence satisfying (i) and (ii) above for every
recursive place selection of this form may be called
{\em Mises-Church stochastic\/}.
Such a selection rule corresponds to a gambling strategy for a
game in which the bits of $A$ are revealed sequentially,
such as by successive coin tosses,  and
a gambler has access only to the past history of the game
when deciding whether to place a bet on the next outcome.
The sequence $A$ is Mises-Church stochastic if
no recursive strategy enables the
gambler to alter the expected gain in the long run.
It is certainly reasonable to accept this as a necessary
condition, however, notice that the definition fails to include
other kinds of strategies (which von Mises clearly had
in mind) such as ``toss a coin, and select
bit $A[n]$ if the toss results in heads,''
which intuitively cannot change the gambler's expected gain
either.  Evidently the crucial point is not that the selection process
is recursive, but that in some sense it has no
information about the sequence $A$; we might say that $A$
must be independent of the selection rule.

The fact that the definition proposed by Church is indeed too
weak to satisfactorily characterize random sequences
is a consequence of a result of
Ville (see \cite{vanLambalgen87a} or \cite{USS90}), i.e.,
there are Mises-Church stochastic sequences which fail to satisfy certain
widely accepted probabilistic laws.  Kolmogorov and (independently)
Loveland (see \cite{USS90}) offered the generalization of a
selection rule given in Definition \ref{place_sel_def} below.
No examples are known of sequences which satisfy (i) and (ii)
for every Kolmogorov-Loveland
selection rule but which fail to satisfy some ``known''
probabilistic law \cite[p.398]{Kolmogorov-Uspenskii87}, though
not every such sequence is algorithmically random in the sense
of Definition \ref{rand_def} \cite{USS90}, \cite{Shen89}.

M. van Lambalgen observed that the relation ``$A$ is algorithmically
random relative to $B$'' could be interpreted as an
{\em independence\/} relation, and that relative randomness
in this sense satisfies the axiomatization of independence
given in \cite{vanLambalgenta}.  In this paper we investigate
the independence properties of algorithmically random sequences
related to notions of subsequence selection.  We consider two kinds
of questions: First, if $A$ is
algorithmically random and $A_0$ is a subsequence
chosen according to a (not necessarily recursive) selection rule,
what conditions must be imposed on the selection rule to guarantee
that $A_0$ remains algorithmically random?  Second,
if $A_1$ denotes the subsequence of {\em nonselected\/} bits,
for what kinds of selection rules are $A_0$ and $A_1$
independent?

The main new result proved here is Theorem
\ref{place_sel_thm}, that $A_0$ and $A_1$ are independent when
$A_0$ is chosen via a {\em bounded\/} Kolmogorov-Loveland
selection rule.  While a Mises-Church place selection
is a special case of a selection rule of this form, the application
we have in mind is that the sequence of queries of
a time-bounded Turing machine may, under certain conditions,
be construed as a bounded Kolmogorov-Loveland place selection.
We indicate, for example, how this fact can be used to
prove the well known result \cite{BG81}
that $\P \neq \NP$ relative to an algorithmically random
oracle.  More recently Kautz and Miltersen
\cite{Kautz-Miltersen94} have used Theorem
\ref{place_sel_thm} to show that relative to a random oracle,
NP does not have {\em measure zero\/} within the exponential
complexity classes $\Eone$ or $\Etwo$,
an (apparently) stronger separation of P from NP.
(``Measure
zero'' refers the resource-bounded measure of
Lutz \cite{Lutz92}.)

\section{Preliminaries}
\label{sect_prelim}

Our notation is for the most part standard.  Unfamiliar
notions from recursion theory can probably be found in
Odifreddi
\cite{Odifreddi89}
or Rogers \cite{Rogers67},
and everything else we need to know can be found in
\cite{Chicken-recipes}.
Let $\NN = \{0,1,2,\ldots\}$ denote the natural numbers.  A
{\em string\/} is an element of $\strings$
or $\cylinders$, where the symbol $\perp$
is called an {\em undefined\/} bit.
The concatenation of strings $x$ and $y$ is denoted $xy$.
For any string $x$,
$|x|$ denotes the length of $x$, and $\lambda$ is the unique string
of length $0$.
If $x \in \cylinders$ and $j,k \in \NN$ with
$0 \leq j \leq k < |x|$, $x[k]$ is the $k$th bit (symbol) of $x$ and
$x[j..k]$ is the string consisting of the $j$th through $k$th
bits of $x$ (note that the ``first'' bit of $x$ is the $0$th).
For an infinite binary sequence $A \in \languages$, the
notations $A[k]$ and $A[j..k]$ are
defined analogously.
For any $x,y \in \cylinders$, $x \sqsubseteq y$
means that if $x[k]$ is defined, then $y[k]$ is also defined
and $x[k] = y[k]$; we say that
$x$ is an {\em initial segment\/}, or {\em predecessor\/},
of $y$ or that $y$ is an {\em extension\/} of $x$.
Likewise for $A \in \languages$,
$x \sqsubseteq A$ means $x[k] = A[k]$ whenever bit $x[k]$
is defined.  Strings $x$ and $y$ are said to be {\em incompatible\/},
or {\em disjoint\/}, if there is no string $z$ which is an extension of
both $x$ and $y$; when $x,y \in \strings$, this simply means that
$x \not\sqsubseteq y$ and $y \not\sqsubseteq x$.

Fix a standard enumeration of $\strings$,
$s_0 = \lambda, s_1 = 0, s_2 = 1, s_3 = 00, s_4 = 01, \ldots$.
A {\em language\/}
is a subset of $\strings$; a language $A$ will be identified with
its characteristic sequence $\chi_A \in \languages$, defined by
$s_y \in A \Iff \chi_A[y] = 1$
for $y \in \NN$.  We will consistently  write
$A$ for $\chi_A$.  $\overline{A}$ denotes the bitwise complement of $A$,
i.e., the set-theoretic complement of the language $A$ in $\strings$.
For $X \subseteq \languages$, $X^c$ denotes the complement of
$X$ in $\languages$.  Since the enumeration of strings above
provides a one-to-one correspondence between strings and
natural numbers, we may also regard an infinite sequence $A$ as
a subset of $\NN$.

Typically  strings in $\cylinders$
will be used to represent {\em partially defined languages}
or partially defined subsets of $\NN$,
and will generally be represented by lower-case greek letters.
For $\sigma \in \cylinders$, when no confusion is likely
to result we will regard $\sigma$, $\sigma\perp^k$, and
$\sigma\perp^{\infty}$ as essentially the same object, since all
specify the same language fragment.  We avoid using the
notation $|\sigma|$ unless $\sigma \in \strings$, however
following \cite{Lutz92} we let $\defined{\sigma}$ denote
the number of {\em defined\/} bits in $\sigma$.
When $\alpha \in \cylinders$ and $\tau \in \strings$, the
notation $\alpha \downarrow \tau$ (``$\tau$ inserted into $\alpha$'')
is defined by
\[
(\alpha \downarrow \tau)[x] = \left\{
\begin{array}{ll} \alpha[x] & \mbox{ if $\alpha[x]$ is defined,} \\[2pt]
\tau[j] & \mbox{ if $x$ is the $j$th undefined }\\
        & \mbox{ position in $\alpha$ and $j < |\tau|$,} \\[2pt]
\perp & \mbox{ otherwise.}
\end{array} \right.
\]
For $A,B \in \languages$, $A/B$ is the subsequence of $A$ {\em selected\/}
by $B$, i.e., if $y_0, y_1, \ldots$ are the positions of the $1$-bits of
$B$ in increasing order, then $(A/B)[x] = A[y_x]$.  Note that $A/B$
is a finite string if $B$ contains only finitely many $1$'s.  For
$\sigma, \tau \in \strings$, $\sigma/\tau$ may be defined analogously.
For $A,B \in \languages$, the sequence $A\oplus B$ is defined by
\[
A\oplus B[x] = \left\{ \begin{array}{ll}
                A[\frac{x}{2}] & \mbox{ if $x$ is even,}\\[4pt]
                B[\frac{x+1}{2}] & \mbox{if $x$ is odd}.
                \end{array}\right.
\]

Let $\langle \cdot, \cdot \rangle : \NN  \times \NN  \into
\NN $ be a fixed recursive bijection.  Then $A^{[i]}$, the
{\em $i$th column of $A$\/}, is defined by
\[
A^{[i]} = \{ n: \langle n,i \rangle \in A \}.
\]
Given any countable sequence \Seq{B}{i} we can define a set
$\bigoplus_i B_i$ whose $i$th column is $B_i$:
\[
\bigoplus_{i}B_i
= \{\langle n,i \rangle : n \in B_i \}.
\]

Let $\varphi_e$  denote the $e$th partial recursive (p.r.)
function, and $\varphi^A_e$  the $e$th p.r.\ function relative
to $A \in \languages$.  We write $\varphi_e(x)\downarrow$ if $\varphi_e$ is
defined on $x$, and $\varphi_e(x)\uparrow$ otherwise; the same
holds for the relativized $\varphi_e^A$.     For $s \in
\NN $,
\[
\varphi_{e,s}(x) = \left\{ \begin{array}{ll}
     \varphi_e(x) & \mbox{if $\varphi_e(x)$ converges in $\leq
s$ steps}\\
\mbox{undefined} & \mbox{otherwise.} \end{array} \right.
\]
$\varphi_e^{\sigma}(x)$ generally abbreviates
$\varphi_{e,\lth{\sigma}}^{\sigma}(x)$; we treat
$\varphi_e^{\sigma}$ or $\varphi_{e,s}$
as a partially defined set or language, i.e., it may be
regarded as an element of $\cylinders$.  As
usual, $W_e = \dom(\varphi_e)$ is the $e$th recursively enumerable
(r.e.) set, $W_{e,s} = \dom(\varphi_{e,s})$, $W_e^A =
\dom(\varphi_e^A)$, and $W_e^{\sigma} =
\dom(\varphi_e^{\sigma})$.

If $\varphi_e^A$ is total, so that $\varphi_e^A = B$ for some
set $B$ (i.e., its characteristic function),
we write $B \Tleq A$; if $B \Tleq A$ and $A \Tleq B$,
we write $A \Teq B$.  $A \Tle B$ means that $A \Tleq B$ but $B
\not \Tleq A$.  The equivalence class $\deg(A) = \{B \in
\languages: A
\Teq B \}$ is called the {\em degree\/} of $A$; $\DD$ denotes
the collection $\{\deg(A): A \in \languages \}$, the {\em Turing
degrees\/} or {\em degrees of unsolvability\/}.   The relation
$\Tleq$ induces a well-defined partial order on $\DD$ (simply
denoted $\leq$) and the operation $\join$ induces a well-defined
least upper bound operation {\boldmath $\cup$} on $\DD$.  The
{\em jump\/} of a set $A$, denoted $A'$, is the set 
\[
\{x: \varphi_x^A(x)\downarrow \},
\]
and $A^{(n)}$ represents the $n$th iterate of the jump of $A$.  
For functions $f:\NN  \into \NN $, by $\deg(f)$ we mean the
degree of the graph of $f$, $\{ \langle x,y \rangle: f(x) = y
\}$.

A string $\sigma \in \cylinders$ defines the subset
$\Ext(\sigma) = \{ A \in \languages : \sigma \sqsubseteq A \}$
of $\languages$, called a {\em cylinder\/}. $\Ext(\sigma)$ is referred
to as an {\em interval\/} if $\sigma \in \strings$; evidently
any cylinder is a union of intervals.
Likewise if
$S$ is a subset of $\cylinders$,
$\Ext(S)$ denotes $\bigcup_{\sigma \in S} \Ext(\sigma)$.
If $S=W_e$ is an r.e.\ set of strings, then $\Ext(S)$ is called
a $\Sigma^0_1$-class; the number $e$ is an {\em index\/} of the
class.  A $\Pi^0_1$-class is the complement of a $\Sigma^0_1$-class.
In general a $\Pi^0_n$-class is the complement of a
$\Sigma^0_n$-class,
and a $\Sigma^0_{n+1}$-class is of the form $\bigcup_i{\cal{T}}_i$,
where $\{ {\cal{T}}_i \}$
is a uniform sequence of $\Pi^0_n$-classes; likewise
a $\Pi^0_{n+1}$-class is of the form $\bigcap_i{\cal{T}}_i$,
where the ${\cal{T}}_i$
are $\Sigma^0_n$-classes.  Here, as elsewhere, a {\em uniform\/} (or
{\em recursive\/}) sequence $\{ {\cal{T}}_i \}$ is one for which there is a
recursive function $f$ such that $f(i)$ is an index for ${\cal{T}}_i$;
an index for the function $f$ may be called an index of the
sequence.  A class is called {\em arithmetical\/} if it is
$\Sigma^0_n$ for some $n$. 

It is also convenient to note that arithmetical classes can be
defined in terms of quantifier complexity.  Let $\LL^*$ be the
language of arithmetic with a set constant $X$ and a membership
symbol $\in$.  Then for a sentence $\phi$ of $\LL^*$ and
$A \in \languages$,
$A \models \phi$ means that $\phi$ is true in the standard
model when $X$ is interpreted as $A$.  A $\Sigma^0_n$-class is of the
form $\{A: A \models \phi \}$,  where $\phi$ is a $\Sigma_n$
sentence of $\LL^*$; likewise for $\Pi^0_n$-classes.  Since
notions of computation can be expressed in a simple way in
$\LL^*$ we can, for example, represent a $\Sigma^0_n$-class in the
form
\[
\{ A : (\exists x_1)(\forall x_2)\ldots (\exists
x_n)[\varphi_e^A(x_1,\ldots, x_n)\downarrow] \}
\]
if $n$ is odd, and in the form
\[
\{ A : (\exists x_1)(\forall x_2)\ldots (\forall
x_n)[\varphi_e^A(x_1,\ldots, x_n)\uparrow] \}
\]
if $n$ is even.  See Rogers \cite{Rogers67} for details.

The definitions of arithmetical classes can all be relativized;
e.g., a $\Sigma^C_1$-class is of the form $\Ext(W_e^C)$, etc.  Note,
for example, that a $\Sigma^{0^{(n-1)}}_1$ -class is an {\em open\/}
$\Sigma^0_n$-class, and a $\Pi^{0^{(n-1)}}_1$-class is a {\em closed\/}
$\Pi^0_n$-class, where we take $\{ \Ext(\sigma) : \sigma \in \strings \}$
as the base of a topology on $\languages$.

By a {\em measure\/} we simply mean a probability distribution
on $\languages$, and for our present purposes it is sufficient
to consider the uniform distribution, i.e., each bit is equally likely to
be a zero or a one, also called Lebesgue measure.
The measure of a subset $\cal{E}$ of $\languages$,
denoted $\Pr(\cal{E})$, can be intuitively
interpreted as the probability that a sequence produced by tossing
a fair coin is in the set $\cal{E}$; in particular
the measure of an interval $\Ext(\sigma)$,
abbreviated $\Pr(\sigma)$, is just $(\frac{1}{2})^{|\sigma|}$
(or $(\frac{1}{2})^{\defined{\sigma}}$ if $\sigma \in \cylinders$).
For $S$ a set of strings,
we abbreviate $\Pr(\Ext(S))$ by $\Pr(S)$;
if $S$ is {\em disjoint\/}, i.e., all strings in $S$ are pairwise
incompatible, then
\[
\Pr(S) = \sum_{\sigma \in S}\Pr(\sigma).
\]
Standard results of measure theory (see \cite{Feller66}) show that
$\cal{E}$ is {\em measurable\/} (meaning that $\Pr(\cal{E})$ is defined)
as long as $\cal{E}$ is a {\em Borel\/} set, i.e., built up from
intervals by some finite iteration of countable union
and complementation operations; in particular arithmetical
classes are always measurable.

\section{Effective Approximations in Measure}
\label{sect-effective} 
  
Equivalent definitions of algorithmic randomness
have been given by Martin-L\"of \cite{MartinLof66},
Levin \cite{Levin73},
Schnorr \cite{Schnorr73}, Chaitin \cite{Chaitin87,Chaitin87b,Chaitin75},
and Solovay \cite{Solovay75}, and generalizations and
closely related variations have been given by
Kurtz \cite{Kurtz81}, Kautz \cite{Kautz91}, and Lutz \cite{Lutz92}.
Here we present the definition due to Martin-L\"of and
the generalization ($n$-randomness) first investigated in \cite{Kurtz81}.
The idea is to characterize a random sequence by
describing the properties of ``nonrandomness'' which it must 
{\em avoid}.  For example, we might examine
successively longer initial segments $\sigma$ of a sequence $A \in
\languages$
and discover that
\begin{eqnarray}
\frac{\mbox{\# of ones in $\sigma$}}{|\sigma|} \geq \frac 3 4 .
\label{lim_cond}
\end{eqnarray}
We would then suspect that $A$ is not a sequence we would
normally think of as random.  A ``test'' for this particular
nonrandomness property can be viewed as a recursive enumeration
of strings $\sigma$ for which (\ref{lim_cond}) holds; if $A$ has
arbitrarily long initial segments satisfying (\ref{lim_cond}),
we reject $A$ as nonrandom.  Now among all possible enumerations
of strings, how do we decide {\em a priori\/} which
ones characterize a
``nonrandomness'' property?  Since ultimately we
expect the nonrandom sequences to form a class with measure
zero, we can require that as we enumerate longer initial
segments in the ``test'', the total measure of their extensions
should become arbitrarily small.  The mathematical content of
this discussion is made precise in following definition.

\begin{definition}\rm
\label{mltestdef}
A {\em Martin-L\"{o}f test} is a recursive sequence of
\s{0}{1}-classes \Seq{\scriptS}{i} with $\Pr(\scriptS_i) \leq 2^{-i}$.  A sequence
$A \in \languages$  is
{\em $1$-random }\ if for every Martin-L\"{o}f test
\Seq{\scriptS}{i}, $A \not\in \I{\scriptS}{i}$.
\end{definition}

The requirement that a test consist of recursively enumerable
sets of strings is a natural starting point but is admittedly
somewhat arbitrary.  A generalized form, where \s{0}{n}-classes
replace \s{0}{1}-classes, first appeared in Kurtz \cite{Kurtz81}.
We will also find it useful to define randomness relative to an oracle.

\begin{definition}\rm
\label{rand_def}
Let $A, C \in \languages$.  $A$
is {\em \s{C}{n}-approximable}, or {\em approximable in
\s{C}{n}-measure}, if there is a recursive sequence
of \s{C}{n}-classes \Seq{\scriptS}{i}   
with $\Pr(\scriptS_{i}) \leq 2^{-i}$ and $A \in
\I{\scriptS}{i}$.  The sequence \Seq{\scriptS}{i} is called
a {\em $\Sigma^C_n$-approximation\/},
or if $n=1$, a {\em Martin-L\"of test\/} or
{\em constructive null cover\/} relative to $C$.
$A$ is {\em $C$-$n$-random }, or {\em $n$-random
relative to $C$}, if $A$ is not
\s{C}{n}-approximable.  If $C$ is recursive then $A$ is {\em
$n$-random}.  We also say $A$ is  {\em
$\omega$-random\/} if $A$ is $n$-random for all $n$.
\end{definition}

The definition of $n$-randomness as given above is difficult to
work with directly when $n>1$,
and there are two basic tools which simplify our work
considerably.  The first is Lemma \ref{opensetlm} below, 
which asserts that $n$-randomness is the same as $1$-randomness 
relative to $0^{(n-1)}$, or more generally, that $(m+n)$-randomness 
relative to an oracle $C$ is the same as $n$-randomness relative 
to $C^{(m)}$.  The second is Theorem \ref{jumpthm}, which shows
that if $A$ is $(n+1)$-random, then $A^{(n)} \Teq A \join 0^{(n)}$. 
These two facts allow many results
on $1$-randomness to be straightforwardly generalized. 
 
The proof of Lemma \ref{opensetlm} depends on the idea that an 
approximation in measure can be replaced by an approximation 
using {\em open\/} classes of the same arithmetical complexity. 
Lemma \ref{apprlm} extends Kurtz' Lemma 2.2a (\cite[p.21]{Kurtz81}) 
with a number of technical refinements which will be needed later.

\begin{lemma} 
\label{measurelm} 
The predicate ``$\Pr(\scriptS) > \epsilon$'' is uniformly \s{C}{n},
where $\scriptS$ is a \s{C}{n}-class and $\epsilon$ is rational.   
Likewise ``$\Pr(\scriptS) < \epsilon$'' is uniformly \s{C}{n} when
$\scriptS$ is a \p{C}{n}-class.   
\end{lemma}  
 
\begin{proof} See Kurtz \cite{Kurtz81}. 
\end{proof} 
 
\begin{lemma} 
\label{apprlm} 
\begin{enumerate.roman} 
 
\item 
For $\scriptS$ a \s{C}{n}-class and $\epsilon > 0$ a rational, we can 
uniformly and recursively obtain the index of a 
\s{C^{(n-1)}}{1}-class\/ (an open \s{C}{n}-class) $\UU \supseteq 
\scriptS$ with $\Pr(\UU) - \Pr(\scriptS) \leq \epsilon$.

\item  
For $\TT$ a \p{C}{n}-class and $\epsilon > 0$ a rational, we can  
uniformly and recursively obtain the index of a  
\p{C^{(n-1)}}{1}-class\/ (a closed \p{C}{n}-class) $\VV \subseteq 
\TT$ with $\Pr(\TT) - \Pr(\VV) \leq \epsilon$.

\item  
For $\scriptS$ a \s{C}{n}-class and $\epsilon > 0$ a rational, we can  
uniformly in $C^{(n)}$ obtain a closed \p{C}{n-1}-class\/ $\VV  
\subseteq \scriptS$ with $\Pr(\scriptS) - \Pr(\VV) \leq \epsilon$.
(If $n \geq 2$, $\VV$ will be a \p{C^{(n-2)}}{1}-class.)
Moreover, if $\Pr(\scriptS)$ is a real recursive in $C^{(n-1)}$, the
index for $\VV$ can be found recursively in $C^{(n-1)}$.  
  
\item  
For $\TT$ a \p{C}{n}-class and $\epsilon > 0$ a rational, we can  
uniformly in $C^{(n)}$ obtain an open \s{C}{n-1}-class\/ $\UU  
\supseteq \TT$ with $\Pr(\UU) - \Pr(\TT) \leq \epsilon$.
(If $n \geq 2$, $\UU$ will be a \s{C^{(n-2)}}{1}-class.)
Moreover, if $\Pr(\TT)$ is a real recursive in $C^{(n-1)}$, the
index for $\UU$ can be found recursively in $C^{(n-1)}$.  
  
\end{enumerate.roman}  
\end{lemma}  

\begin{proof}
\end{proof}
 
The characterization in terms of approximation by {\em open\/} sets  
now follows easily from Lemma \ref{apprlm}. 
 
\begin{lemma}  
\label{opensetlm}  
Let $A, C \in \languages$, $n \geq 1$, and $m \geq 0$.  Then
$A$ is \s{C^{(m)}}{n}-approximable $\Iff$ $A$ is   
\s{C}{m+n}-approximable.  
\end{lemma} 
 
\begin{proof} 
($\Rightarrow$) Immediate, since any \s{C^{(m)}}{n}-class is a 
\s{C}{m+n}-class. 
 
\noindent ($\Leftarrow$) We show that a 
\s{C}{m+1}-approximation can be replaced by a 
\s{C^{(m)}}{1}-approximation; the result will then follow by 
induction on $n$.  Let $C \in \languages$ and $m \geq 1$ be 
arbitrary.  Suppose \Seq{\scriptS}{i} is a \s{C}{m+1}-approximation. 
By Lemma \ref{apprlm}(i) we can uniformly find for each $i$ a 
\s{C^{(m)}}{1}-class $\UU_i \supseteq \scriptS_i$ with 
$\Pr(\UU_i) - \Pr(\scriptS_i) \leq 2^{-i}$.  Thus $\Pr(\UU_{i+1})
\leq 2^{-i}$, so $\{\UU_{i+1}\}$ is a
\s{C^{(m)}}{1}-approximation and $\I{\scriptS}{i} \subseteq 
\bigcap_{i}\UU_{i+1}$. 
\end{proof} 
 
We conclude this section by stating several results we will need which 
have been known for $1$-randomness, and which can be easily 
generalized using Lemma \ref{opensetlm}.  The first of these 
is known as the ``Universal Martin-L\"{o}f Test''.  For a proof 
see \cite{MartinLof66} or \cite{Kautz91}. 
 
\begin{theorem}[Martin-L\"{o}f] 
\label{univthm} 
For any $C \in \languages$ and any $n \geq 1$ there exists a 
universal \s{C}{n}-approximation.  That is, there is a recursive sequence 
of \s{C^{(n-1)}}{1}-classes \Seq{\UU}{i}, with $\Pr(\UU_i) \leq
2^{-i}$, such that every \s{C}{n}-approximable set is in \I{\UU}{i}.  
\end{theorem} 
 
The next result is a characterization of 
$n$-randomness due to Solovay.  It shows that the conditions 
``$\Pr(\scriptS_i) \leq 2^{-i}$'' and ``$A \in \I{\scriptS}{i}$'' in
definition \ref{rand_def} are both stronger than necessary.  A proof
can be found in \cite{Chaitin87} or \cite{Kautz91}.
 
\begin{theorem}[Solovay] 
\label{solovaythm} 
Let $A$, $C \in \languages$  and $n \geq 1$.
$A$ is $C$-$n$-random $\Iff$ for 
every recursive sequence of \s{C}{n}-classes \Seq{\scriptS}{i} with 
$\sum_{i}\Pr(\scriptS_i) < \infty$, $A$ is in only finitely many
$\scriptS_i$.  
\end{theorem}

We close this section with one of the single most useful facts
about $n$-randomness, since it allows many results about
\s{0}{1}-approximations to be extended to
\s{0}{n}-ap\-prox\-i\-ma\-tions.  It strengthens
a result originally due to Sacks that the class
$\{ A : A' \Teq A \oplus 0' \}$ has measure one \cite{Stillwell72}.
The proof is in \cite{Kautz91}.

\begin{theorem}
\label{jumpthm}
For $n \geq 0$, if $A$ is $(n+1)$-random, then
$A^{(n)} \Teq A \oplus 0^{(n)}$.
\end{theorem}

\section{Independence and place selections}

The definition below describes
a very general selection process which encompasses several special
cases of interest. The process may be pictured as follows,
as suggested in \cite{USS90}:
Suppose the sequence $A$ is represented as a row of cards
laid face down; on the face of the $i$th card is either a
zero or a one, corresponding to $A[i]$.  We have two 
functions, $F$ and $G$, which are used to select some of the cards
to create a second sequence, which we continue to call a ``subsequence''
even though the order of the cards may be changed.
Both $F$ and $G$ look at the
history of the selection process, that is, the sequence
of cards turned over so
far.  The value of $F$ is a natural number indicating the position
of the next card to be turned over.  The value of $G$ is either 0
or 1; if the value is 0, the card is merely turned over and observed,
while if the value is 1 the card is also {\em selected\/}, i.e., added
onto the end of the subsequence.
 
\begin{definition}\rm
\label{place_sel_def}
A {\em Kolmogorov-Loveland place selection\/} \cite{USS90}
is a pair of
partial recursive functions $F: \strings \rightarrow \NN$ and
$G: \strings \rightarrow \{0,1 \}$ (possibly relative to
an oracle $C$).
Let $A \in \languages$; $F$ and $G$ select a subsequence $Q^*$
from $A$ as follows.
First define sequences of strings
$\xi_0 \sqsubseteq \xi_1 \sqsubseteq \cdots$ and
$\rho_0 \sqsubseteq \rho_1 \sqsubseteq \cdots$ such that
$\xi_0 = \rho_0 = \lambda$,
$\xi_{j+1} = \xi_jA[F(\xi_j)]$, and
$\rho_{j+1} = \rho_jG(\xi_j)$
(with the proviso that $\xi_{j+1}$ is undefined if $F(\xi_j) = F(\xi_i)$
for some $i < j$ or if either $F$ or $G$ fails to converge).  If
$\xi_j$ and $\rho_j$ are defined for all $j$ let
$Q = \lim_j \xi_j$  and $R = \lim_j \rho_j$.
Thus $Q$ represents the sequence of all bits of $A$ examined by $F$,
in the order examined.  A given bit $Q[j] = A[F(\xi_j)]$
is included in the subsequence $Q^*$ just if $G(\xi_j) = 1$, i.e.\
$F$ determines which bits of $A$ to examine, and
$G$ determines which ones to include in the sequence $Q^*$.
Formally we define $Q^* = Q/R$.  A 
Kolmogorov-Loveland place selection will be called {\em bounded\/}
if the function $G$ is determined by a partial
recursive function $H: \strings \rightarrow \NN$
(possibly relative to oracle $C$)
with the following properties:
\begin{enumerate.roman}
\item
$H$ is nondecreasing, i.e., if $\xi \sqsubseteq \xi'$ then
$H(\xi) \leq H(\xi')$,
\item
$H$ is unbounded, i.e., if $\xi_j$ and $\rho_j$ are defined for all
$j$ then $\lim_j H(\xi_j) = \infty$, and
\item
$G$ is determined by $H$ according to the rule
\begin{eqnarray*}
F(\xi) < H(\xi) &\Rightarrow& G(\xi) = 0 \\
F(\xi) \geq H(\xi) &\Rightarrow& G(\xi) = 1.
\end{eqnarray*}
\end{enumerate.roman}
It is also useful to define a sequence $B$ by $B[z] = 1$ if and
only if for some $j$, $F(\xi_j) = z$ and $G(\xi_j) = 1$, so that
$N = A/\overline{B}$ consists of the ``nonselected'' bits of $A$,
in their natural order.
\end{definition}
 
If it is always the case that $F(\xi) = n$, where $n=|\xi|$, then
we have a Mises-Church place selection.  Note that if in addition
$G(\xi)$ depends only on the length of $\xi$, then the selected
subsequence is of the form $A/B$, for a fixed sequence $B$.

In general it is not difficult to show that a subsequence
selected from an $n$-random sequence is also
$n$-random.

\begin{theorem}
\label{subseqthm}
Let $A,B,C \in \languages$; suppose $F$ and $G$ are partial
recursive in $B$ and determine a K-L place selection.  If
$A$ is algorithmically random relative to $B\oplus C$, then
$Q^*$ is algorithmically random relative to $C$.
\end{theorem}

\begin{proofof}{Sketch of proof: }
Suppose $\{\UU_i\}$ is a
$\Sigma^C_1$-approximation of $Q^*$; we can construct
a $\Sigma^{B \oplus C}_1$-approximation $\{ \scriptS_i \}$ of $A$.
Let $\UU_i = \Ext(W^B_e)$.  For $\sigma$ enumerated in $W^B_e$,
start with a string $\alpha = \perp^{\infty}$.
simulate $F$ and $G$:  If $G(\xi) = 1$, let $\alpha[F(\xi)] =
\mbox{ next bit of $\sigma$}$.  If $G(\xi) = 0$, then split into
two strings, let $\alpha_0[F(\xi)] = 0$ and $\alpha_1[F(\xi)] = 1$.
When no more bits of $\sigma$ are available, enumerate
the corresponding string $\alpha$ into $S_i$.  Total measure of all
strings $\alpha$ associated with $\sigma$ is not more than
$\Pr(\sigma)$, since each time we split into two strings an
additional bit of $\alpha$ is defined.
\end{proofof}

Note that it follows, using Lemma \ref{opensetlm} and Theorem
\ref{jumpthm} and taking $C = 0^{(n-1)}$, that if $A$ is
$n$-random relative to $B$, then $Q^*$ is $n$-random.  We can also
conclude that if $A$ is $n$-random relative to $B$ (and $B$ is infinite)
then $A/B$ is $n$-random.
Note in particular that, as predicted by von Mises' intuition,
the place selection function, or the
sequence $B$, does not have to be recursive.  In fact, it is
a consequence of the following result that if $A$ is $n$-random, then
{\em almost every\/} place selection preserves the randomness
properties of $A$.

\begin{theorem}
\label{measure_thm}
Let $A,C \in \languages$.  If
$\{ B: \mbox{ $A$ is $\Sigma^{B \oplus C}_1$ approximable} \}$
has positive measure, then $A$ is $\Sigma^C_1$-approximable.
\end{theorem}

It then follows,
that if $A$ is $n$-random,
\begin{eqnarray}
\label{B_set}
\{ B: \mbox{ $A$ is $n$-random relative to $B$} \}
\end{eqnarray}
has measure one:  note that by
Lemma \ref{opensetlm} and Theorem \ref{jumpthm}, $A$ is
\s{B}{n}-approximable iff it is \s{B^{(n-1)}}{1}-approximable,
and $B^{(n-1)} \Teq B \oplus 0^{(n-1)}$.  Hence if the set
(\ref{B_set}) has measure $< 1$, then
$\{ B: \mbox{ $A$ is \s{B \oplus 0^{(n-1)}}{1}-approximable} \}$
has positive measure, which would imply that $A$ is
\s{0^{(n-1)}}{1}-approximable, i.e., not $n$-random.


Proving that a selected subsequence is actually {\em independent\/} of
the nonselected bits is quite a bit more subtle.  We begin with
the relatively straightforward version below.  Although it is actually
a consequence of Theorem \ref{place_sel_thm}, the idea of the
proof is much more in evidence in this simpler setting. 

\begin{theorem}
\label{indep_thm} 
Let $A, B, C \in \languages$.
If $A \oplus B$ is $1$-random relative to $C$, then
$A$ is $1$-random relative to $B\oplus C$. 
\end{theorem}

\begin{proof} 
As the relativization is straightforward we will suppress the
oracle $C$ for readability.  We show that
if $A$ is \s{B}{1}-approximable, then $A \join B$ is
\s{0}{1}-approximable.
Suppose $A$ is \s{B}{1}-approximable,
say by \Seq{\TT}{i}.  Let $f$ be a recursive function giving the
indices of the classes $\TT_i$, i.e., $\TT_i = \Ext(W^B_{f(i)})$.
Fix $i$ and let $e = f(i)$.  We describe a uniform procedure
for enumerating a set of strings $S_i$ such that 
$\{\Ext(S_j)\}$ is a \s{0}{1}-approximation of
$A \join B$.  Let $S_{i,s}$ be the set of strings
\[
\{ \sigma \join \tau : \lth{\sigma} = \lth{\tau} = s 
\mbox{ \rm \& } (\exists \sigma' \subset \sigma)(\exists \tau' \subset
\tau)
[\sigma' \in W_e^{\tau'} \mbox{ \rm \& } \Pr(\Ext(W_e^{\tau'})) \leq
2^{-i}]
\}
\]
and $S_i = \Union_s S_{i,s}$.  Note that $S_i$ is r.e.  
Certainly $A \join B$ is in $\Ext(S_i)$, since some initial segment
$\sigma'$ of $A$ is in $W_e^B$ and hence in $W_e^{\tau}$ for some
$\tau \subset B$ with $s = \lth{\tau} \geq \lth{\sigma'}$.  Thus 
$\sigma \join \tau$ is enumerated in $S_{i,s}$, where $\sigma$
is the initial segment of $A$ of length $s$.

To show that $\Pr(\Ext(S_i)) \leq 2^{-i}$, since $\Ext(S_{i,s})
\subseteq \Ext(S_{i,s+1})$ it will suffice to show that for each
$s$, $\Pr(\Ext(S_{i,s})) \leq 2^{-i}$.  Fix $s$ and fix a
string $\tau$ of length $s$.  Let $\tau^*$ be the longest initial
segment of $\tau$ such that $\Pr(\Ext(W_e^{\tau^*})) \leq 2^{-i}$.
Then for every string of the form $\sigma \join \tau$ in $S_{i,s}$ 
there must be some $\sigma' \subset \sigma$ in $W_e^{\tau^*}$, so
the measure contributed to $S_{i,s}$ by strings of the form 
$\sigma \join \tau$ cannot exceed
\begin{eqnarray*}
&&\sum_{\sigma' \in W_e^{\tau^*}}2^{-\lth{\sigma'}}\cdot
2^{-\lth{\tau}}\\
&=& 2^{-\lth{\tau}} \cdot \Pr(\Ext(W_e^{\tau^*}))\\
&\leq & 2^{-s}\cdot 2^{-i}
\end{eqnarray*}
(where we have tacitly assumed that $W_e^{\tau^*}$ is disjoint).
There are $2^s$ strings $\tau$ of length $s$, so the total measure of
$\Ext(S_{i,s})$ is at most $2^{-i}$.
\end{proof}

It follows that if $A \oplus B$ is $n$-random, then
$A$ is $n$-random relative to $B$.
If $A$ is \s{B}{n}-approximable, then $A$ is
\s{B^{(n-1)}}{1}-approximable (by Lemma \ref{opensetlm}), and hence is  
\s{B\join 0^{(n-1)}}{1}-approximable by Theorem \ref{jumpthm} (note
that $B$ is $n$-random by Theorem \ref{subseqthm}). Thus $A \join B$ is
\s{0^{(n-1)}}{1}-approximable by Theorem \ref{indep_thm}
i.e., not $n$-random.

A converse has been proved by van Lambalgen \cite{vanLambalgenta}:
if $B$ is $n$-random and $A$ is $n$-random relative to $B$,
then $A \oplus B$ is $n$-random; then it follows that
$B$ is also $n$-random relative to $A$.


The result below encompasses Theorem \ref{indep_thm}
as well as a number of other cases of interest, such as
Mises-Church place selections and selections of the form
$A/B$ for a fixed $B$.
The remainder of this section will be
devoted to a proof of the result.

\begin{theorem}
\label{place_sel_thm}
Let $A,B,C \in \languages$.
Let $F$ and $H$ be
partial recursive functions relative to $B$ which
determine a bounded Kolmogorov-Loveland
place selection, 
and let $N$ and $Q^*$ be as in Definition \ref{place_sel_def}.
If $A$ is algorithmically
random relative to $B \oplus C$ and $N$ is infinite,
then $N$ is algorithmically random relative to $Q^* \oplus C$.
\end{theorem}

Notice that using Lemma \ref{opensetlm} and Theorem
\ref{jumpthm} and taking $C = 0^{(n-1)}$ and $B$
recursive, it follows from Theorem \ref{place_sel_thm} that if $A$ is
$n$-random then $N$ is $n$-random relative to $Q^*$.
Moreover, by the converse to Theorem \ref{indep_thm} noted above,
it follows that $Q^*$ is also $n$-random relative
to $N$.

Before beginning the proof, we isolate a counting argument that
will be needed.

\begin{lemma}
\label{counting_lemma}
Let $s \in \NN$ and let $f$ be a real-valued function on $\strings$
whose domain includes all strings of length at most $s$.
Suppose there is a constant $c$ such that for every string $\tau_0$
of length $s$,
\[
\sum_{\tau \sqsubseteq \tau_0} f(\tau) \leq c.
\]
Then
\[
\sum_{j=0}^s \sum_{|\tau| = j} 2^{-|\tau|} f(\tau) \leq c.
\]
That is, the domain of $f$ can be viewed as a full binary
tree of height $s$ with a real value assigned to each node;
if the sum along each branch is bounded by $c$,
then the sum, for $j \leq s$,  of the average value of the nodes at
level $j$ is also bounded by $c$.
\end{lemma}
\begin{proof}
First note that if $M$ is any $n \times m$ real matrix and $c$ is
a constant such that the sum of each column is bounded by $c$,
then the total of the averages of the rows is bounded by $c$.
This is easy to see, since if
\[
\sum_{j=0}^{n-1} M(j,k) \leq c
\]
for each $k=0, \ldots, m-1$, then
\begin{eqnarray}
\label{M_inequality}
\sum_{j=0}^{n-1} \left(\frac{1}{m}\sum_{k=0}^{m-1} M(j,k)\right) =
\frac{1}{m} \sum_{k=0}^{m-1} \sum_{j=0}^{n-1} M(j,k)
\leq \frac{1}{m}\sum_{k=0}^{m-1} c = c.
\end{eqnarray}
Let $m=2^s$ and $n=s+1$ and let $\tau_0, \tau_1, \ldots, \tau_{m-1}$ be
a list of all strings of length $s$.  Define an $n \times m$ matrix $M$
by
\[
M(j,k) = \alt{f(\tau_k[0..j-1])}{j>0}
             {f(\lambda)}{j=0}.
\]
That is, the values along the branch $\tau_k$ correspond to
the $k$th column of $M$, so the sums of the columns of $M$ are
bounded by $c$.  For each $j = 0,\ldots,n-1$, the values in row
$j$ are of the form $f(\tau)$ for strings $\tau$ of length $j$;
there are $2^j$ possible values $f(\tau)$,
each of which must occur $2^{s-j}$ times
in row $j$. Hence
\[
\sum_{k=0}^{m-1}M(j,k) = \sum_{|\tau| = j}2^{s-j} f(\tau)
\]
and so
\begin{eqnarray*}
\sum_{j=0}^s \sum_{|\tau| = j} 2^{-|\tau|}f(\tau)
 &=& \sum_{j=0}^{n-1} \sum_{|\tau| = j}2^{-j}f(\tau) \\
 &=& \sum_{j=0}^{n-1} \sum_{|\tau| = j}\frac{1}{2^s}2^{s-j}f(\tau) \\
 &=& \sum_{j=0}^{n-1} \frac{1}{2^s}\sum_{k=0}^{m-1}M(j,k) \\
 &\leq& c
\end{eqnarray*}
by (\ref{M_inequality}).
\end{proof}


\begin{proofof}{Proof of Theorem \protect{\ref{place_sel_thm}}}
Suppose that $\{U_i \}$ is a constructive null cover relative to
$Q^*$ with $N \in \bigcap_i \Ext(U_i)$.
We suppress the oracle $C$.
We will exhibit a
constructive null cover $\{S_i \}$
(relative to $B$)
of $A$ by describing a
procedure for enumerating, uniformly in $B$, a set
$S_i$ such that $A \in \Ext(S_i)$ and $\Pr(S_i) \leq 2^{-i}$.
Fix $i$; we can uniformly obtain an index $e$ for which
$U_i = W_e^{Q^*}$.  Let $e$ be fixed throughout the remainder of
the proof.

The enumeration of $S_i$ may be loosely described in the following
way: we ``guess'' an initial segment $\tau \sqsubseteq Q^*$ and
enumerate strings $\sigma \in W_{e,t}^{\tau}$ for 
$t = 0, 1, \ldots, |\tau|$
as long as $\Pr(W_{e,t}^{\tau}) \leq 2^{-i}$.  For each pair
$(\sigma, \tau)$ we attempt to ``reconstruct'' an initial segment
$\alpha \sqsubseteq A$.
That is, we attempt to construct a string $\alpha \in \cylinders$
such that when the place selection is applied to $\alpha$
(stopping when $F$ attempts to examine an undefined bit), the
selected
subsequence is exactly $\tau$ and the nonselected subsequence
is exactly $\sigma$.  $S_i$ then consists of an enumeration of
all the strings
$\alpha$ obtained for all possible ``guesses'' $\tau$.

We give below a formal definition of a partial recursive function
\[
S: \strings \times \strings \rightarrow \cylinders
\]
which produces $\alpha$ from an input pair $(\sigma, \tau)$.
The construction of $S$ can be informally described as follows:
During the course of the construction,
we define sequences
$\hat{\xi}_0 \sqsubseteq \hat{\xi}_1 \sqsubseteq \cdots$ and
$\alpha_0 \sqsubseteq \alpha_1 \sqsubseteq \cdots$ of strings.
$S$ will be attempting to simulate the action of $F$ and $H$ on the
sequence $A$, and each string
$\hat{\xi}_i$ represents  what $S$ believes to be the
corresponding actual value of the string
$\xi_i$ of Definition \ref{place_sel_def}.  The strings
$\alpha_i$ represent approximations to $A$.
The idea is that $S$ will regard $\tau$ as an initial
segment of $Q^*$ and $\sigma$ as an initial segment of $N$;
as $S$ reads through $\tau$ it uses the value of $F(\hat{\xi}_i)$
to determine
the original position that each of the selected bits
originally had in $A$, and
in effect what $S$ does is replace
each bit of $\tau$ in its proper position in $\alpha_i$.
To simulate the action that $F$ and $H$ would have taken on
$A$, $S$ needs the correct value of the bit examined by $F$
at each step.
There are only two possibilities at any stage $i$
in the simulation:  if
$F(\hat{\xi}_i) \geq H(\hat{\xi}_i)$, the correct value
should be the next bit of $\tau$.  If $F(\hat{\xi}_i) 
< H(\hat{\xi}_i)$, the positions in $\alpha_i$ to the left of
$H(\hat{\xi}_i)$ which have not yet been defined using 
bits from $\tau$ must correspond to bits of $N$, so they
can be filled in with the initial part of
$\sigma$ and the correct value of $A$ can be determined from
$\alpha_i \downarrow \sigma$.  Obviously if
$\tau \not\sqsubseteq Q^*$ or $\sigma \not\sqsubseteq N$, the simulation
will be incorrect and if $S(\sigma,\tau)$ converges to some value
$\alpha$, it is unlikely that $\alpha \sqsubseteq A$.  The fact
that we only consider pairs $(\sigma,\tau)$ for which
$\sigma \in W_{e,t}^{\tau}$ and $\Pr(W_{e,t}^{\tau}) \leq 2^{-i}$
will be used to ensure that $\Pr(S_i) \leq 2^{-i}$.

{\bf Construction of $S$:}  At stage $0$, let
\[
\alpha_0 = \perp^{\infty},\ \  \hat{\xi}_0 = \lambda, \ \  
\mbox{ and } t_0 = 0,
\]
where $t_i$ is a marker indicating the next unexamined bit of
$\tau$.

At stage $k+1$, let $h = H(\hat{\xi}_k)$ and let $u$ be the
number of undefined bits in $\alpha_k[0..h-1]$.
If $|\sigma| < u$, then the construction diverges and
$S(\sigma,\tau)$ is undefined.
If $|\sigma| \geq u$ let
$\alpha_k^* = \alpha_k \downarrow \sigma[0..u-1]$.
If $|\sigma| = u$ and $|\tau| = t_k$, then the construction
terminates at stage $k+1$ with value 
\[
S(\sigma,\tau) = \alpha_k^*.
\]
Otherwise one of the following cases applies:

\begin{description}
\item[Case 1:] $F(\hat{\xi}_k) < h$.  Then let
\begin{eqnarray*}
\hat{\xi}_{k+1} &=& \hat{\xi}_k \alpha_k^*[F(\hat{\xi}_k)], \\
\ t_{k+1} &=& t_k, \\
\mbox{and }\ \alpha_{k+1} &=& \alpha_k.
\end{eqnarray*}

\item[Case 2:] $F(\hat{\xi}_k) \geq h$.  If $t_k \geq |\tau|$
then the construction diverges and $S(\sigma,\tau)$ is 
undefined.  Otherwise let
\begin{eqnarray*}
\hat{\xi}_{k+1} &=& \hat{\xi}_k \tau[t_k], \\
t_{k+1} &=& t_k + 1, \\
\mbox{and for all $z\in \NN$,  }
\alpha_{k+1}[z] &=& \left\{ \begin{array}{ll}
     \tau[t_k] & \mbox{ if $z = F(\hat{\xi}_k)$} \\
     \alpha_k[z] & \mbox{ otherwise.}
     \end{array} \right. 
\end{eqnarray*}
\end{description}

The crucial properties of the function $S$ are summarized in
the claim below.
The proof of Claim \ref{S_claim} is technical, and we 
will postpone giving the details until the end of this section. 

\begin{claim}
\label{S_claim}
Let $A$, $Q^*$, and $N$ be as in Definition \ref{place_sel_def}.
\begin{enumerate.roman}
\item
If $S(\sigma,\tau)$ converges with value $\alpha$ and
$S(\sigma', \tau')$ converges with value $\alpha'$, then $\sigma' \sqsubseteq
\sigma$ and $\tau' \sqsubseteq \tau$ imply that $\alpha' \sqsubseteq
\alpha$.
\item
If $S(\sigma,\tau)$ converges and $\sigma'$ is a proper initial
segment of $\sigma$, there is no
proper extension $\tau'$ of $\tau$ for which $S(\sigma', \tau')$
converges.
\item
If $S(\sigma, \tau)$ converges with value $\alpha$, then
$\Pr(\alpha) = 2^{-|\sigma|-|\tau|}$.
\item
If $\tau \sqsubseteq Q^*$ and $\sigma \sqsubseteq N$ and
$S(\sigma, \tau)$ converges with value $\alpha$, then
$\alpha \sqsubseteq A$.
\item
For any strings $\tau' \sqsubseteq Q^*$ and $\sigma' \sqsubseteq N$,
there exist $\sigma, \tau \in \strings$ such that $\tau' \sqsubseteq
\tau \sqsubseteq Q^*$, $\sigma' \sqsubseteq \sigma \sqsubseteq N$,
and $S(\sigma, \tau)$ converges.
\end{enumerate.roman}
\end{claim}

We next show how to obtain a constructive null cover $\{S_i\}$ of
$A$, given the properties listed above of the function $S$.
First define a recursive function $t$ by
\[
t(\tau) = \max \{r \leq |\tau| :
                \Pr(W^{\tau}_{e,r}) \leq 2^{-i} \}.
\]
Then for $s \in \NN$ and $\tau \in \strings$, let
\begin{eqnarray*}
S_{i,\tau} &=& \{ S(\sigma, \tau) :
 \mbox{ For some $\sigma' \sqsubseteq \sigma$,
  $\sigma' \in W_{e,t(\tau)}^{\tau}$}  \} \\
S_{i,s} &=& \bigcup_{|\tau| \leq s} S_{i,\tau} \\
S_i &=& \bigcup_s S_{i,s}
\end{eqnarray*}

We need to verify the following two facts.
\begin{claim}
\label{A_claim}
$A \in \Ext(S_i)$.
\end{claim}
\begin{proofof}{Proof of Claim \protect{\ref{A_claim}}}
\ Since by assumption
$N \in \Ext(U_i)$, there is an initial segment $\sigma' \sqsubseteq N$
with $\sigma' \in W_e^{Q^*}$.  It follows that for some
$\tau' \sqsubseteq Q^*$, $\sigma' \in W_{e,|\tau'|}^{\tau'}$.
By Claim \ref{S_claim}(iv) and (v), there are strings $\sigma, \tau$ such
that $\sigma' \sqsubseteq \sigma \sqsubseteq N$, $\tau' \sqsubseteq \tau
\sqsubseteq Q^*$, and $S(\sigma, \tau) = \alpha \sqsubseteq A$.
Since $\Pr(W_{e,|\tau|}^{\tau}) \leq \Pr(U_i) \leq 2^{-i}$, it follows
that $\alpha$ is enumerated into $S_{i,\tau}$ and hence into $S_i$.
\end{proofof}

\begin{claim}
\label{measure_claim}
$\Pr(S_i) \leq 2^{-i}$.
\end{claim}
\begin{proofof}{Proof of Claim \protect{\ref{measure_claim}}}
Since it is clear that $S_{i,s} \subseteq S_{i,s+1}$ for all $s$,
it suffices to fix $s$ and show that $\Pr(S_{i,s}) \leq 2^{-i}$.
For each $\tau$, $|\tau| \leq s$, define a set
\[
B(\tau) = \{ \sigma : \mbox{ $S(\sigma, \tau)$ converges and
     some $\sigma' \sqsubseteq \sigma$ is in $W_{e,t(\tau)}^{\tau}$} \}.
\]
Note that $S_{i,\tau} = \{S(\sigma,\tau) : \sigma \in B(\tau) \}$
and that
\[
S_{i,s} =
\{S(\sigma, \tau) :
   \mbox{ $|\tau| \leq s$ and $\sigma \in B(\tau)$} \}. 
\]
Define a string $\sigma \in B(\tau)$ to be an {\em initial\/} 
string
if there is no proper prefix $\sigma' \sqsubseteq \sigma$
such that $\sigma' \in B(\tau')$ for some 
$\tau' \sqsubseteq \tau$.  Let
$B^*(\tau)$ denote the initial strings in $B(\tau)$.

Consider a string $\tau$, $|\tau| \leq s$.  If $\sigma \in B(\tau)$,
there is a unique shortest predecessor $\sigma' \sqsubseteq \sigma$
such that $S(\sigma', \tau')$ converges for some $\tau'$
comparable to $\tau$.
By Claim \ref{S_claim}(ii), we may assume that
$\tau' \sqsubseteq \tau$. It follows  that 
$\sigma'$ is an initial string, and
if $\sigma' \neq \sigma$ then $\sigma$ is not an initial string.
Then $\bigcup \{B^*(\tau') : \tau' \sqsubseteq \tau \}$ is
prefixfree, and so
\begin{eqnarray}
\label{sum_of_paths}
\sum_{\tau' \sqsubseteq \tau}\Pr(B^*(\tau'))
= \Pr\left(\bigcup \{B^*(\tau') : \tau' \sqsubseteq \tau \}\right)
\leq \Pr(W_{e,t(\tau)}^{\tau}) \leq 2^{-i}.
\end{eqnarray}
It further follows from
Claim \ref{S_claim}(i) that 
\[
\Ext\{S(\sigma, \tau) :
         \mbox{ $|\tau| \leq s$ and $\sigma \in B(\tau)$} \} 
=  \Ext\{S(\sigma, \tau) :
         \mbox{ $|\tau| \leq s$ and $\sigma \in B^*(\tau)$} \}.
\]
Then
\begin{eqnarray*}
\Pr(S_{i,s}) &=&  \Pr\{S(\sigma, \tau) :
   \mbox{ $|\tau| \leq s$ and $\sigma \in B^*(\tau)$} \} \\
&\leq& \sum_{\stackrel{ |\tau| \leq s}{\sigma \in B^*(\tau)}}
        \Pr(S(\sigma,\tau)) \\
&=& \sum_{\stackrel{ |\tau| \leq s}{\sigma \in B^*(\tau)}}
        2^{-|\sigma| - |\tau|} 
        \mbox{ \ \ by Claim \protect{\ref{S_claim}}(iii) }   \\
&=& \sum_{j=0}^s \sum_{|\tau| = j} 2^{-|\tau|}\Pr(B^*(\tau)) \\
&\leq& 2^{-i},
\end{eqnarray*}
where the last inequality follows from 
(\ref{sum_of_paths}), using Lemma \ref{counting_lemma}
with $f(\tau) = \Pr(B^*(\tau))$.
\end{proofof}

All that remains is to 
verify the properties listed in Claim \ref{S_claim}.

{\em Proof of Claim \protect{\ref{S_claim}}.}\ Parts 
(i) and (ii) are straightforward consequences of the fact that in
the construction of $S$, the inputs are used from left to right.
Suppose $S(\sigma,\tau)$ converges to $\alpha$ at 
stage $k+1$, and let $\alpha_k$, $\hat{\xi}_k$, and $t_k$ be
the final values of the variables in the construction; suppose
also that  $S(\sigma',\tau')$ converges to $\alpha'$ at
stage $j+1$ and let $\alpha_j$, $\hat{\xi}_j$, and $t_j$
be the corresponding final values. 
Let $h_j = H(\hat{\xi}_j)$ and $h_k = H(\hat{\xi}_k)$.
Suppose $\sigma' \sqsubseteq \sigma$ and 
$\tau' \sqsubseteq \tau$.  Since
the inputs are used from left to right, the construction
of $S(\sigma,\tau)$ is identical to that of $S(\sigma',\tau')$
up to the stage at which the latter converges, i.e., 
$\alpha_j \sqsubseteq \alpha_k$ and $\hat{\xi}_j \sqsubseteq
\hat{\xi}_k$.
Since $H$ is nondecreasing, and no bits of $\alpha_j$ to the left 
of $h_j$ can be defined after stage $j+1$ in the construction, we know
$\alpha_j[0..h_j-1] = \alpha_k[0..h_j-1]$. Since the number of undefined
bits in $\alpha_j[0..h_j-1]$ is exactly $|\sigma'|$ (by the
definition of the convergence of $S$), we have
\[
\alpha' = (\alpha_j \downarrow \sigma') \sqsubseteq (\alpha_k
\downarrow \sigma) = \alpha,
\]
which establishes (i).  

For (ii), we use the notation of the preceding
paragraph, and assume that $\sigma'$ is a {\em proper\/}
initial segment of $\sigma$ and that $\tau \sqsubseteq \tau'$.
Again the construction of $S(\sigma,\tau)$ must be identical to
that of $S(\sigma', \tau')$ up to the stage at which one of them
converges.  Suppose that $S(\sigma', \tau')$ converges first, i.e.,
that $k < j$.  Then $\alpha_k \sqsubseteq \alpha_j$,
$\hat{\xi}_k \sqsubseteq \hat{\xi}_j$, $h_k \leq h_j$, and
$\alpha_k[0..h_k-1] = \alpha_j[0..h_k-1]$, implying that
\begin{eqnarray*}
|\sigma| &=&
\mbox{ \# of undefined bits in $\alpha_k[0..h_k-1]$} \\
 &\geq& \mbox{ \# of undefined bits in $\alpha_j[0..h_j-1]$} \\
 &=& |\sigma'|,
\end{eqnarray*}
contradicting the fact that $\sigma'$ is a proper initial segment
of $\sigma$.  Therefore $j \leq k$, and so $t_j \leq t_k$ and
$|\tau'| \leq |\tau|$, i.e., $\tau'$ cannot be a proper extension
of $\tau$.


For part (iii), note that when the computation $S(\sigma,\tau)$ 
converges at a stage $k+1$, exactly $|\tau|$ bits of $\alpha_k$
have been defined, so the value $\alpha = \alpha_k \downarrow \sigma$
has exactly $|\tau| + |\sigma|$ defined bits.

Parts (iv) and (v) involve slightly more work.  We first 
prove inductively that the following conditions (a)--(e) hold.
Suppose that $\sigma \sqsubseteq N$ and $\tau \sqsubseteq Q^*$
and $S(\sigma,\tau)$ converges.  Then 
(using the notation of Definition \ref{place_sel_def}) 
at each nonterminating stage $k$ in
the computation of $S(\sigma,\tau)$,
\begin{description}
\item (a) $|\xi_k/\rho_k| = t_k$,
\item (b) $\hat{\xi}_k = \xi_k$,
\item (c) $\alpha_k \sqsubseteq A$,
\item (d) for all $z < H(\hat{\xi}_k)$, 
$\mbox{$\alpha_k[z]$ is defined\ } \Longleftrightarrow
 B[z] = 1$, and
\item (e) $\alpha_k^* \sqsubseteq A$.
\end{description}

{\bf Base step:}
\ When $k = 0$, (a), (b), and (c) are immediate.  For (d), note that
no bit of $A$ to the left of $H(\lambda)$
is ever selected, so $B[z]=0$ for $z < H(\lambda)$; also
$\alpha_0[z]$ is undefined for all $z$.  For (e), let $h = H(\lambda)$;
since $\alpha_0$ is undefined at all bits, 
\[
\alpha_0^* = (\alpha_0 \downarrow \sigma[0..h-1]) = \sigma[0..h-1]
= N[0..h-1] = A[0..h-1],
\]
where the last equality follows from the fact that $B[z] = 0$ for
all $z < h$.

{\bf Induction step.}
\ Suppose (a)--(e) hold for $j \leq k$.
For (a), (b),
and (c) there are two cases:
\begin{description}
\item[Case 1:]\ $F(\hat{\xi}_k) < H(\hat{\xi}_k)$.
\begin{description}
\item (a)\ By construction $t_k = t_{k+1}$, and
since $\hat{\xi}_k = \xi_k$ by the induction hypothesis,
$F(\xi_k) <  H(\xi_k)$ and so $\rho_k+1 = \rho_k0$.  Hence 
\[
|\xi_{k+1}/\rho_{k+1}| = |\xi_k/\rho_k| = t_k = t_{k+1}.
\]
\item (b)\ By the induction hypothesis, 
$\alpha_k^* \sqsubseteq A$ and $\hat{\xi}_k = \xi_k$;
in view of the latter it is enough to show that the
$(k+1)$st bits of $\hat{\xi}_{k+1}$ and $\xi_{k+1}$ are equal:
\[
\hat{\xi}_{k+1}[k] = \alpha_k^*[F(\hat{\xi}_k)] 
= A[F(\xi_k)] = \xi_{k+1}[k].
\]
\item (c) By construction $\alpha_{k+1} = \alpha_k \sqsubseteq A$.
\end{description}
\item[Case 2:]\ $F(\hat{\xi}_k) \geq H(\hat{\xi}_k)$.
\begin{description}
\item\ (a) By construction $t_k = t_k +1$, and
since $\hat{\xi}_k = \xi_k$ by the induction hypothesis,
$F(\xi_k) \geq  H(\xi_k)$, so $\rho_k+1 = \rho_k1$;  hence 
\[
|\xi_{k+1}/\rho_{k+1}| 
= |\xi_k/\rho_k| + 1 = t_k + 1 = t_{k+1}.
\]
\item\ (b)  Again $\hat{\xi}_k = \xi_k$ by the induction hypothesis,
and since 
$\xi_{k+1}/\rho_{k+1} \sqsubseteq \tau \sqsubseteq Q^*$,
\begin{eqnarray}
\label{tau-equality}
\hat{\xi}_{k+1}[k] = \tau[t_k] = (\xi_{k+1}/\rho_{k+1})[t_k] =
\xi_{k+1}[k].
\end{eqnarray}
\item\ (c) We have $\alpha_k \sqsubseteq A$ by the induction 
hypothesis, and by construction $\alpha_{k+1}[z] = \alpha_k[z]$
for all $z \neq F(\hat{\xi}_k)$.  By (\ref{tau-equality}) above,
$\tau[t_k] = \xi_{k+1}[k]$, which is equal to $A[F(\xi_k)]$.   Then
\[
\alpha_{k+1}[F(\hat{\xi}_k)] = \tau[t_k] = A[F(\xi_k)] = A[F(\hat{\xi}_k)],
\]
so $\alpha_{k+1} \sqsubseteq A$.
\end{description}
\end{description}

For (d), suppose that $z < H(\hat{\xi}_{k+1})$.
Note that $H(\hat{\xi}_{k+1}) = H(\xi_{k+1})$ by (b) above.
Then 
\begin{eqnarray*}
\mbox{$\alpha_{k+1}[z]$ is defined } &\Longleftrightarrow&
    \mbox{ for some $j \leq k$, $z = F(\hat{\xi}_j) \geq H(\hat{\xi}_j)$} \\
&\Longleftrightarrow&
    \mbox{ for some $j \leq k$, $z = F(\xi_j) \geq H(\xi_j)$} \\
&\Longleftrightarrow& B[z] = 1.
\end{eqnarray*}


Part (e) follows from (c) and (d): 
Let $h = H(\hat{\xi}_{k+1})$, $\beta = B[0..h-1]$, and let 
$u$ denote the number of undefined bits in $\alpha_{k+1}[0..h-1]$. 
Then
\begin{eqnarray*}
\alpha_{k+1}^*/\beta &=& \alpha_{k+1}/\beta \mbox{ \ by (d)} \\
&=& A/\beta \mbox{ \ by (c)},\\
\mbox{so \ }\alpha_{k+1}^*/\overline{\beta} &=& \sigma[0.. u-1] \\
&=& A/\overline{\beta}
\mbox{ \ since $\sigma \sqsubseteq N = A/\overline{B}$.}
\end{eqnarray*}
But $\alpha_{k+1}^*/\beta = A/\beta$ and 
$\alpha_{k+1}^*/\overline{\beta} = A/\overline{\beta}$ imply that
$\alpha_{k+1}^*[0..h-1] = A[0..h-1]$.  Since $\alpha_{k+1}^*[z]
= \alpha_{k+1}[z]$ for all $z \geq h$, we have $\alpha_{k+1}^*
\sqsubseteq A$ by (c).
This completes the induction.

Now Claim \ref{S_claim}(iv) follows from part (e),
since when $S(\sigma,\tau)$ converges at
a stage $k+1$, the final value is $\alpha = \alpha^*_{k} \sqsubseteq A$.

The final step is to verify part (v) of Claim \ref{S_claim}.
Consider any integer $k$ for which $F(\xi_k) \geq H(\xi_k)$.
Let $h = H(\xi_k)$, and let
\begin{eqnarray}
\label{tau_def}
\tau &=& \xi_k/\rho_k \\
\label{sigma_def}
\mbox{and \ \ } \sigma &=& A[0..h-1]/\overline{B}.
\end{eqnarray}
Clearly $\tau \sqsubseteq Q^*$ and $\sigma \sqsubseteq N$.
It is also the case that
$S(\sigma,\tau)$ converges:  at stage $k+1$ in the construction,
we have $t_k = |\xi_k/\rho_k| = |\tau|$ 
by part (a), $h = H(\hat{\xi}_k) = H(\xi_k)$ by (b), and by (d),
\begin{eqnarray*}
\mbox{\# of undefined bits in $\alpha_k[0..h-1]$} &=&
     \mbox{\# of zeros in $B[0..h-1]$} \\
&=& |\sigma| \mbox{\hspace{2em} by (\protect{\ref{sigma_def}}),}
\end{eqnarray*}
so $S(\sigma,\tau)$ converges.
Since $Q^*$ and $N$ are assumed to be infinite,
it is always possible to find a $k$ such that
$F(\xi_k) \geq H(\xi_k)$ and such that the strings
$\tau$ and $\sigma$ of (\ref{tau_def}) and (\ref{sigma_def}) are
as long as desired; therefore $\sigma$ and $\tau$ can be
found which extend any given 
$\tau' \sqsubseteq Q^*$ and $\sigma' \sqsubseteq N$.

The proof of Claim \ref{S_claim}, and hence the proof of 
Theorem \ref{place_sel_thm}, is now complete.
Thank you for your support.
\end{proofof}


\end{document}